\documentclass[journal]{IEEEtran}
\usepackage{xcolor}
\usepackage{amsmath}
\usepackage{amsfonts}
\usepackage{mathtools,amssymb,bm}
\usepackage{epstopdf}
\usepackage{hyperref}
\usepackage[footnotesize]{subfigure}
\newcommand{\RN}[1]{%
  \textup{\uppercase\expandafter{\romannumeral#1}}%
}

\usepackage{cite}

\ifCLASSINFOpdf
\else
\fi
\usepackage{amsmath}
\usepackage{algorithmic}

\usepackage{array}
\usepackage{fixltx2e}

\usepackage{stfloats}
\usepackage{url}

\begin{document}

\title{Signal Reconstruction using Blind Super-resolution with Arbitrary Sampling}
%
%
%

\author{Hoomaan~Hezaveh,
        Milad~Javadzadeh,
        and~Mohammad Hossein~Kahaei}
\maketitle

\begin{abstract}
In this paper the problem of blind super-resolution of sparse signals using arbitrary sampling scheme and atomic lift is discussed. After comprehensive description on blind super-resolution problem, it is shown that using Prolate Spheroidal Wave Functions (PSWFs), it is possible to derive a new Semi-Definite Program (SDP) for the blind super-resolution problem. Unlike the previous results, the newly proposed SDP can localize spikes without magnitude recovery. Several numerical simulations were conducted to compare the performance of the proposed method with the recent related research.
\end{abstract}


%
\IEEEpeerreviewmaketitle

\section{Introduction}
%
%
%
%
\setlength{\parindent}{1em}\IEEEPARstart{M}{any} applications consider the problem of recovering a signal ${x(t)}$ convolved with a function ${g(t)}$ known as Point Spread Function (PSF)\cite{chi2016guaranteed}. Without any assumption on ${g(t)}$ and ${x(t)}$ the problem is ill posed. Therefore, many methods have been proposed based on different assumptions \cite{ahmed2013blind,wang2016blind}. In practice, it is common to assume that the signal of interest is sparse and can be modeled with a sum of spikes\cite{}. Though sparsity is an important assumption for signal recovery, it is not enough to solve the blind deconvolution problem successfully. Another important assumption is on PSF. Inspired by \cite{ahmed2013blind}, it is possible to assume that the PSF lies in a known subspace but its orientation in the subspace remains unknown. This assumption provides us a mathematical model for PSF which is previously used in blind deconvolution problem \cite{chi2016guaranteed}.\par
\setlength{\parindent}{1em}Conventional sparsity based algorithms, which are known as blind compressed sensing methods, assume that frequencies lie on a predefined grid which is originated from the dictionary\cite{Chi2011}. However, in many applications true frequencies can take any value and are not confined to the points on a grid. Therefore, basis mismatch problem arises\cite{Chi2011}. Cand\'{e}s et.al introduced a new attitude towards sparse recovery which yielded sparse grid-less methods\cite{Candes2014}. These methods achieved higher precision in frequency estimation and signal recovery than grid-based methods provided that a minimum frequency separation condition is satisfied. However, most of the recent sparse grid-less methods assume a grid based sampling scheme. Mahata and Hyder developed a sparse arbitrary sampling based frequency estimation method to gain higher precision in recovery of signals which are bandlimited and confined to a time window using Prolate Spheroidal Wave Functions (PSWFs)\cite{mahata2016frequency}.
\par
\setlength{\parindent}{1em}The problem of blind deconvolution of sparse signals in gridless compressed sensing regime via lifting was previously investigated\cite{chi2016guaranteed}. In this paper we analyze the same problem with different sampling scheme and therefore different mathematical perspective. We arbitrarily sample the received signal and extend the mathematical theory of \cite{mahata2016frequency} to fit in blind gridless compressed sensing. Using the fact that the spikes of a sparse time signal are confined to a time window and the received signal is bandlimited due to PSF, we use PSWFs to express the variables of an optimization problem. This presentation leads to a new Semi-Definite Programming (SDP) to solve blind grid-less compressed sensing problem. Numerical results show that the proposed method achieves higher rate of successful spike recovery than recent methods.\par

\setlength{\parindent}{1em} The rest of the paper is as follow: In section \RN{2} we formulate the blind sparse gridless compressed sensing problem. In section \RN{3} we derive a SDP for the latter problem using arbitrary sampling scheme, PSWFs and atomic norm minimization. In section \RN{4} numerical results are presented to evaluate the performance of the new method.

\section{Problem Formulation}
\setlength{\parindent}{1em} Consider signal ${x(t)}$ consisting of ${K}$ Dirac functions at positions ${\bar{\tau}_k\in [-\frac{T_{max}}{2},\frac{T_{max}}{2}]}$ with amplitudes ${a_k\in\mathbb{C}}$, where ${T_{max}}$ is signal duration in time. Therefore, we have
\begin{equation}\label{1}
  x(t)=\sum_{k=1}^{K}a_k\delta(t-\bar{\tau}_k)
\end{equation}
where ${\delta(.)}$ is Dirac function. According to the fact that received signal is the result of convolution ${x(t)\ast g(t)}$, we have
\begin{equation}\label{2}
  y(t)=x(t)\ast g(t)=\sum_{k=1}^{K}a_kg(t-\bar{\tau}_k)
\end{equation}
where ${g(t)}$ is the PSF. By taking the Fourier transform of (\ref{2}) we have
\begin{equation}\label{3}
  Y(f)=X(f)G(f)=\big(\sum_{k=1}^{K}a_ke^{-j2\pi f\bar{\tau}_k}\big)G(f)
\end{equation}
where ${X(f)}$ and ${G(f)}$ are Fourier transform of ${x(t)}$ and ${g(t)}$ respectively and ${j=\sqrt{-1}}$. Note that ${G(f)}$ is bandlimited to ${[\frac{-B_{max}}{2},\frac{B_{max}}{2}]}$. By sampling ${Y(f)}$ in arbitrary points ${f_m\in[\frac{-B_{max}}{2},\frac{B_{max}}{2}] , m=1,\dots,M}$ we can write
\begin{equation}\label{4}
  Y(f_m)=\big(\sum_{k=1}^{K}a_ke^{-j2\pi f_m\bar{\tau}_k}\big)G(f_m),\:m=1,...,M.
\end{equation}
Consider ${\bar{f}_m=f_mT_{max}}$ and ${\tau_k=\bar{\tau}_k/T_{max}}$. Therefore,
\begin{equation}\label{5}
  Y(\bar{f}_m)=\big(\sum_{k=1}^{K}a_ke^{-j2\pi \bar{f}_m\tau_k}\big)G(\bar{f}_m),\:m=1,...,M.
\end{equation}
We can rewrite (\ref{5}) in a compact form as
\begin{equation}\label{6}
  \boldsymbol y=diag(\boldsymbol g)\boldsymbol x
\end{equation}
where ${\boldsymbol y\in \mathbb{C}^{M\times 1}}$ is the observation vector, ${diag(\boldsymbol g)}$ presents an ${M\times M}$ digonal matrix with ${\boldsymbol g}$ as its main diagonal and ${\boldsymbol x\in \mathbb{C}^{M\times 1}}$ is the Fourier transform of signal ${x(t)}$ at arbitrary sampled points ${\bar{f}_m}$.\par
\setlength{\parindent}{1em}In this paper, we assume that ${\boldsymbol g=\boldsymbol S\boldsymbol h}$ lies in a known low dimensional subspace ${\boldsymbol S\in \mathbb{C}^{M\times L}}$ with unknown orientation ${\boldsymbol h\in \mathbb{C}^L}$ and ${L\ll M}$. This assumption is not far from the reality and holds in many applications\cite{chi2016guaranteed}. Consider the matrix ${\boldsymbol S^T=[\boldsymbol s_0,...,\boldsymbol s_M]}$ where ${\boldsymbol s_m\in C^L}$ is the ${m}$-th column of the matrix ${\boldsymbol S^T}$. By defining ${x_m}$ and ${y_m}$ as the ${m}$-th elements of ${\boldsymbol x}$ and ${\boldsymbol y}$, we can use the lifting trick and write(\ref{6}) as \cite{ahmed2013blind}
\begin{eqnarray*}
  y_m &=& \boldsymbol s^{T}_m \boldsymbol h x_m=\boldsymbol s^{T}_m \boldsymbol h \boldsymbol e^{T}_m \boldsymbol x \\
  &=&\boldsymbol e^{T}_m(\boldsymbol x\boldsymbol h^T)\boldsymbol s_m:=\boldsymbol e^{T}_m\boldsymbol Z^{\ast} \boldsymbol s_m=\langle\boldsymbol Z^{\ast},\boldsymbol e_m\boldsymbol s^{H}_m\rangle.\\
\end{eqnarray*}
where ${e_m}$ is the ${m-th}$ standard basis vector of ${\mathbb{R}^M}$. Therefore, we can consider ${\Upsilon:C^{M\times L}\rightarrow C^M}$ such that
\begin{equation}\label{7}
  \boldsymbol y=\Upsilon(\boldsymbol Z)
\end{equation}
where $${\boldsymbol Z=\boldsymbol x \boldsymbol h^T=\sum_{k=1}^{K}a_k\boldsymbol c(\tau_k)\boldsymbol h^T\in C^{M\times L}}$$ and $${\boldsymbol c(\tau_k)=[e^{-j2\pi f_1\tau_k},e^{-j2\pi f_2\tau_k},\ldots,e^{-j2\pi f_M\tau_k}]^T}$$ for ${\tau_k\in [-\frac{1}{2},\frac{1}{2}]}$. ${\boldsymbol x}$ and ${\boldsymbol g}$ can be recovered up to a scaling factor using left and right singular vectors of ${\boldsymbol Z}$. Hence, now the problem is to recover ${\boldsymbol Z^*}$ from ${\boldsymbol y}$. This problem is ill-posed because the number of unknowns is much more than the number of measurements. It is clear that ${\boldsymbol Z}$ columns are composed of ${K}$ complex sinusoids with the same frequencies. We use this joint spectral sparsity of ${\boldsymbol Z}$ columns to recover ${\boldsymbol Z^*}$ from ${\boldsymbol y}$ by minimizing the atomic norm for joint spectrally sparse signals\cite{chi2016guaranteed}.

\section{Atomic Norm Minimization and Arbitrary Sampling}
\setlength{\parindent}{1em}To recover ${\boldsymbol Z^*}$ from ${\boldsymbol y}$, one should define the respective atomic set ${\mathcal{A}}$ as
$${\mathcal{A}=\{\boldsymbol A(\tau,\boldsymbol u)=\boldsymbol c(\tau)\boldsymbol u^H \in C^{M\times L}|\tau \in [-\frac{1}{2},\frac{1}{2}],\|\boldsymbol u\|_2=1\}}.$$
Atomic norm minimization is the problem of finding minimum number of atoms in ${\mathcal{A}}$ to define ${\boldsymbol Z^*}$ and can be expressed as \cite{chi2014joint}
\begin{align}\label{8}
  \|\boldsymbol Z\|_{\mathcal{A}} & = inf\{t>0\: :\boldsymbol Z\in tConv(\mathcal{A})\} \\
   & =\underset{\boldsymbol u_k\in C^L:\|\boldsymbol u_k\|_2=1}{\underset {\tau_k\in[-\frac{1}{2},\frac{1}{2}]}{inf}}\{\sum_{k}c_k|\boldsymbol Z=\sum_{k}c_k \boldsymbol A(\tau_k,\boldsymbol u_k),c_k\geq0\}\nonumber
\end{align}
where ${Conv(\mathcal{A})}$ denotes the convex hull of the set ${\mathcal{A}}$. Therefore, we should solve the following problem
\begin{equation}\label{9}
  \underset{\boldsymbol Z}{min} \|\boldsymbol Z\|_{\mathcal{A}}\:\:\: s.t. \:\:\:\boldsymbol y=\Upsilon(\boldsymbol Z)
\end{equation}
where ${\boldsymbol y\in C^M}$ is the measurement vector and ${\Upsilon}$ is previously defined. There are many SDP characterizations of atomic norm minimization problem for different scenarios\cite{chi2014joint,mahata2016frequency,tang2013compressed}. Most of the previous SDPs consider a uniform grid-based sampling scheme\cite{tang2013compressed,chi2014joint}. As presented by \cite{mahata2016frequency}, it is possible to derive a SDP characterization of atomic norm minimization problem based on arbitrary sampling scheme and PSWFs, which can provide us an orthogonal basis on ${\mathbb{L}^2}$ as the set of all square integrable functions on ${[-1/2,1/2]}$. We present the following lemma to cast (\ref{9}) in a tractable manner.
\newtheorem{Lemma}{Lemma}
\begin{Lemma}\label{lemma1}
Consider ${\mathbb{H}}$ to be the set of all ${M\times M}$ Hermitian matrices and ${\boldsymbol P:[-\frac{1}{2},\frac{1}{2}]\rightarrow{\mathbb {H}}}$ be such that ${\boldsymbol P(\tau)=\boldsymbol A(\tau,\boldsymbol u)\boldsymbol A^\ast(\tau,\boldsymbol u)}$. Let ${\mathbb{K}}$ be closed conic hull of the set ${\{\boldsymbol P(\tau)\in \mathbb{H}:\tau \in [-\frac{1}{2},\frac{1}{2}]\}.}$ Then,
\begin{align}\label{10}
  \|\boldsymbol Z\|_{\mathcal{A}} = &\underset{\boldsymbol W\in \mathbb{R}^{L\times L},\boldsymbol Q \in \mathbb{K}}{ arg\:min } \frac{1}{2}(Tr(\boldsymbol W)+\boldsymbol e^\ast \boldsymbol Q \boldsymbol e)\nonumber \\
   &\:\:\:\:\:\: s.t. \left[
            \begin{array}{cc}
              \boldsymbol W & \boldsymbol Z^\ast \\
              \boldsymbol Z & \boldsymbol Q \\
            \end{array}
          \right]\succeq0.
\end{align}
Also, if ${\boldsymbol Z=\sum_{k}\hat{c}_k\boldsymbol A(\hat{\tau}_k,\hat{\boldsymbol u}_k)}$ is such that ${\|\boldsymbol Z\|_{\mathcal{A}}=\sum_{k}\hat{c}_k}$, then ${\boldsymbol W=\sum_{k}\hat{c}_k\hat{\boldsymbol u}_k\hat{\boldsymbol u}_k^\ast}$, ${\boldsymbol Q=\sum_{k}\hat{c}_k\boldsymbol P(\hat{\tau}_k)}$ is a solution to (\ref{10}). Conversely if, ${\boldsymbol Q=\sum_{k}\hat{c}_k\boldsymbol P(\hat{\tau}_k)}$ is a solution to (\ref{10}) for some ${\hat{c}_k>0}$, then the corresponding optimum value of ${\boldsymbol W}$ is ${\sum_{k}\hat{c}_k\hat{\boldsymbol u}_k\hat{\boldsymbol u}_k^\ast}$, ${\boldsymbol Q=\sum_{k}\hat{c}_k\boldsymbol P(\hat{\tau}_k)}$, therefore ${Tr(\boldsymbol W)=\sum_{k}\hat{c}_k=\|\boldsymbol Z\|_{\mathcal{A}}}$ and ${\boldsymbol Z=\sum_{k}\hat{c}_k\boldsymbol c(\hat{\tau}_k) \hat{\boldsymbol u}_k^\ast.}$
\end{Lemma}

\par
\setlength{\parindent}{1em}The proof of \textit{Lemma 1} appears in appendix A.
To present a tractable algorithm solving (\ref{10}), a finite parametrization of ${\mathbb{K}}$ is needed. For ${\boldsymbol Q=\sum_{k}\hat{a}_k\boldsymbol P(\hat{\tau}_k)}$ as a solution to (\ref{10}), it is crucial that ${\boldsymbol Q\in \mathbb{K}}$. According to the fact that our signal is limited to a time window and also band-limited because of PSF, it is possible to use PSWFs as basis to express ${\boldsymbol Q\in \mathbb{K}}$. First, a brief description on PSWFs is given and then we use \textit{Lemma 2} from \cite{mahata2016frequency} to characterize ${\boldsymbol Q\in \mathbb{K}}$. Consider $${c=\pi\boldsymbol B_{max},}$$and let ${\mathbb{L}^2}$ be as previously defined. For any ${r\in \mathbb{L}^2}$, PSWFs are the eigenfunctions of the linear map ${\xi : \mathbb{L}^2 \rightarrow \mathbb{L}^2}$ such that
\begin{equation}\label{10_1}
  (\xi r)(f)=\int^1_{-1}e^{jc\zeta f}r(\zeta)d\zeta, \:\:\:\:\:\:\:\forall f\in [-1,1],
\end{equation}
and therefore, the PSWF ${\varphi_j}$ should satisfy $${\xi\varphi_j=\lambda_j\varphi_j}$$ where ${\lambda_j}$ is the ${j-th}$ eigenvalue of ${\xi}$. Note that ${\varphi_j}$s are produced according to ${|\lambda_j|}$ where ${|.|}$ denotes the absolute value operator. Specifically, ${|\lambda_j|}$s reduce to zero when ${j}$ increases beyond ${2c/\pi}$. This means that we can limit the number of needed PSWFs according to our working precision. One can find ${d}$ by ${d=\{min\:\:\:\: \frac{j}{2}\:|\:\:\:\:|\lambda_j|<\epsilon\}}$ where ${\epsilon}$ is the working precision. More information on PSWFs can be found in \cite{xiao2001prolate}.

\begin{Lemma}[\cite{mahata2016frequency}]
${\boldsymbol Q\in \mathbb{K}}$ if and only if there exists ${v_0\in \mathbb{R}}$ and ${v_k\in \mathbb{C},\: k=1,\dots,d}$ such that ${\boldsymbol T\succeq0,\Psi(\boldsymbol T)\succeq0}$ satisfying $${\boldsymbol Q_{jl}=\boldsymbol h_{jl}^T\Phi^{-1}[v_d^\ast,\dots,v_1^\ast,v_0,v_1,\dots,v_d]^T,}$$ where
\begin{align}
  \boldsymbol h_{jl}(k)&= \varphi_{k-1}((f_j-f_l)/B_{max}),\boldsymbol \Phi_{kj}= \varphi_{j-1}((k-d-1)/d),\nonumber \\
  \boldsymbol T:&= toep(v_0,v_1,\dots,v_d),\boldsymbol J_1=[\boldsymbol I_d\: \boldsymbol 0_{d\times1}],\boldsymbol J_2=[\boldsymbol 0_{d\times1},\boldsymbol I_d].\nonumber \\
  \Psi(\boldsymbol T):&= \tan^2(\theta_0/2)(\boldsymbol J_1+\boldsymbol J_2)\boldsymbol T(\boldsymbol J_1+\boldsymbol J_2)^\ast \nonumber \\
   &\:\:\:\:\:\: -(\boldsymbol J_1-\boldsymbol J_2)\boldsymbol T(\boldsymbol J_1-\boldsymbol J_2)^\ast,\nonumber
\end{align}
Also, ${\theta_0=c/d,c=\pi B_{max}}$, ${\boldsymbol I_d}$ is a ${d\times d}$ identity matrix and ${\boldsymbol 0_{d\times1}}$ is a ${d\times 1}$ zero vector.
In addition, if there exists a singular measure$${\mu(\tau)=\sum_{k=1}^{K}c_ku(\tau-\tau_k),K\leq d,|c_k|>0}$$
with ${u(\tau)}$ as the unit step function, satisfying $${\boldsymbol Q=\int_{-\frac{1}{2}}^{\frac{1}{2}}\boldsymbol P(\tau)d\mu(\tau),}$$ then it is unique and ${\boldsymbol T}$ is singular with the Vandermonde decomposition
\begin{equation}\label{10_2}
 \boldsymbol T=\sum_{k=1}^{K}c_k\omega(2\pi\tau_k B_{max}/d)\omega^\ast(2\pi\tau_k B_{max}/d),
\end{equation}
where ${\omega(\theta):=[1,e^{-j\theta},\ldots,e^{-jd\theta}]^{\ast}}$.
\end{Lemma}
Using \textit{Lemma 1} and \textit{Lemma 2}, a new algorithm to solve (\ref{9}) is obtained as
\begin{align}\label{11}
  \underset{\boldsymbol W\in \mathbb{R}^{L\times L},v_0\in\mathbb{R},v_1,\dots,v_d\in \mathbb{C}} {minimize} & \:\:\:\:\frac{1}{2}(Tr(\boldsymbol W)+\boldsymbol e^\ast\boldsymbol Q\boldsymbol e) \\
  s.t. & \:\:\:\:\left[
           \begin{array}{cc}
             \boldsymbol W & \boldsymbol Z^\ast \\
             \boldsymbol Z & \boldsymbol Q \\
           \end{array}
         \right]\succeq0,\boldsymbol y = \Upsilon(\boldsymbol Z),
   \nonumber\\
   \boldsymbol Q_{jl}=&\boldsymbol h_{jl}^T\Phi^{-1}[v_d^\ast,\dots,v_1^\ast,v_0,v_1,\dots,v_d]^T, \nonumber\\
   \boldsymbol T:=& toep(v_0,v_1,\dots,v_d), \nonumber\\
    \Psi(\boldsymbol T):= &\tan^2(\theta_0/2)(\boldsymbol J_1+\boldsymbol J_2)\boldsymbol T(\boldsymbol J_1+\boldsymbol J_2)^\ast \nonumber \\
     & -(\boldsymbol J_1-\boldsymbol J_2)\bold T(\boldsymbol J_1-\boldsymbol J_2)^\ast\succeq0.\nonumber
\end{align}
According to the fact that ${T}$ has a toeplitz structure and is singular with rank ${K<d+1}$, one can recover ${\{\bar{\tau}_k\}_1^K}$ via Proney's method\cite{yang2015gridless}. Having ${\{\bar{\tau}_k\}}$s, ${\boldsymbol x}$ and ${\boldsymbol g}$ are estimated up to a scaling factor using left and right singular vectors of ${\boldsymbol Z}$. In the case of noisy measurements, one should add ${\frac{1}{\gamma}\|\hat{\boldsymbol y}-\boldsymbol \Upsilon(\boldsymbol Z)\|_2^2}$ to the cost function of (\ref{11}) and delete ${\boldsymbol y=\Upsilon(\boldsymbol Z)}$. In this case, ${\hat{\boldsymbol y}=\boldsymbol y + \boldsymbol n}$ where ${\boldsymbol n}$ is generated using ${\mathcal{C}\mathcal{N}(0,\sigma^2\boldsymbol I)}$, ${\gamma=\sigma\sqrt{Mln(B_{max})}}$ \cite{yang2015gridless} and ${\boldsymbol Z}$ is added to the optimization variables. Note that our method does not utilize dual problem to localize spikes. However, the authors verify that spike recovery using dual polynomial derived from (\ref{11}) is possible. According to the fact that sampling is not conducted on a grid, we are not dealing with orthogonal Fourier basis. Thus, dual certificate proofs proposed before \cite{chi2016guaranteed} are not applicable anymore. This problem is due to the squared Fej\'{e}r kernel's high leakage when arbitrary sampling scheme is applied. Finding an appropriate kernel is left for future studies.

\section{Numerical Results}
To evaluate the performance of the proposed method numerical simulations were conducted. In these simulations we show that the proposed approach has a higher rate of successful spike recovery than the recent method of \cite{chi2016guaranteed}. The simulations are divided to three sections. First, we show how the rate of successful spike recovery is affected by varying the number of spikes (${K}$). In sub-section \ref{subDS}, the dimension of the unknown PSF (${L}$) and in sub-section \ref{subSLN} Signal to Noise Ratio (SNR) is changed to investigate the behaviour of the rate of successful spike recovery under different circumstances. In order to sample arbitrarily, ${f_1=0,f_M=B_{max}=64}$ and each ${f_m,m=1,\ldots,M}$ is obtained uniformly from the interval ${(0,64)}$ while ${M<B_{max}}$. Spikes are separated by at least ${1/M}$. The subspace ${\boldsymbol S\in \mathbb{R}^{M\times L}}$ and the vector ${\boldsymbol h \in \mathbb{R}^{L}}$ are drawn from i.i.d standard Gaussian distribution. Considering ${\hat{\boldsymbol Z}}$ as the estimate of ${\boldsymbol Z}$, normalized mean square error (NMSE) ${\frac{\|\hat{\boldsymbol Z}-\boldsymbol Z^{\ast}\|_F^2}{\|\boldsymbol Z^{\ast}\|^2_F}}$ is used to compare the results. Also, mean square error (MSE) is used to compare spike localization accuracy. All of the results are compared to the state of the art \cite{chi2016guaranteed}. The simulations were conducted in MATLAB using CVX toolbox\cite{grant2014cvx}. In each part 100 Monte-Carlo simulations were conducted.
\subsection{Number of the Spikes}\label{subNS}
\begin{figure}[!t]
  \centering
  \includegraphics[scale=0.301]{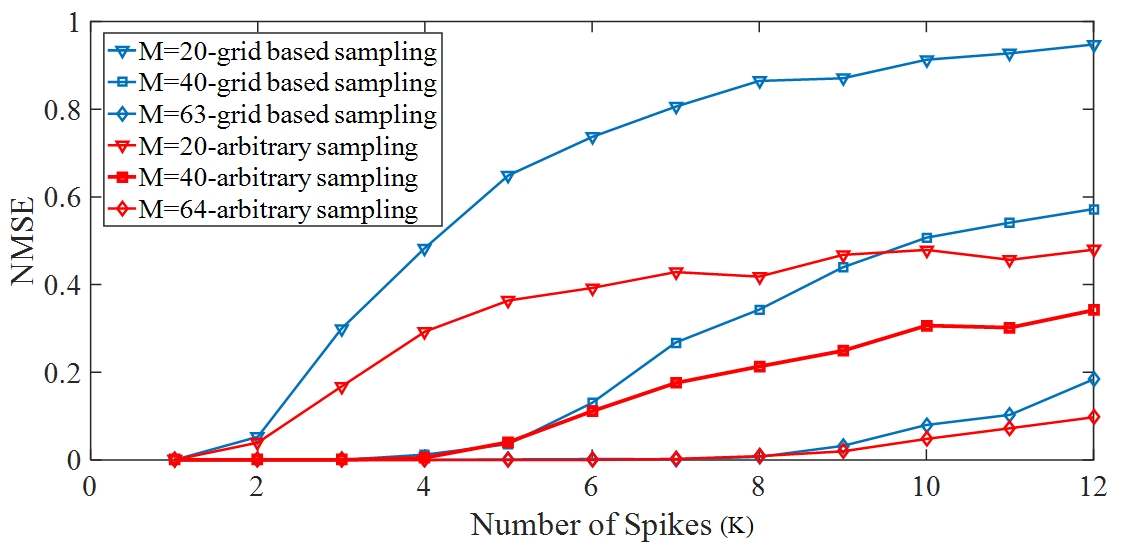}
  \caption{NMSE comparison of the proposed method (arbitrary sampling) with the method of \cite{chi2016guaranteed} for various ${K}$'s (grid-based sampling). }\label{figure1}
\end{figure}

\begin{figure}[!t]
  \centering
  \includegraphics[scale=0.16]{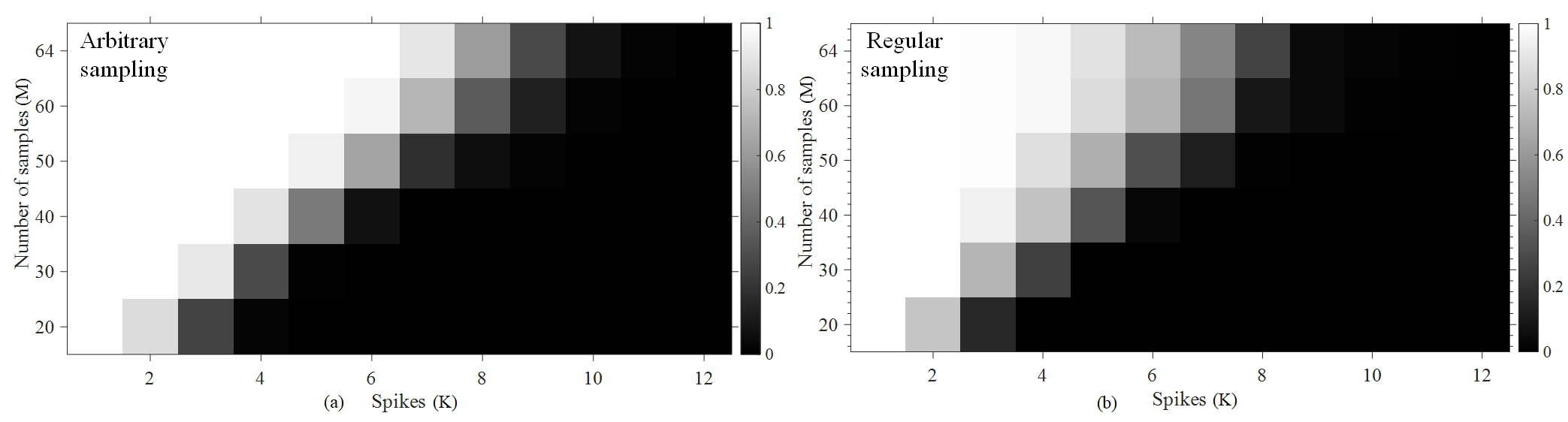}
  \caption{Phase transition diagrams for various ${K}$'s and ${M}$'s. a) The proposed method b) The method of \cite{chi2016guaranteed} }\label{figure2}
\end{figure}

In this part, the number of the spikes is varied from ${1 \text{ to } 12}$. The dimension of the PSF subspace is set to (${L=3}$). The number of the measurements is varied from 20 to 63. Figure \ref{figure1} depicts the NMSE of matrix ${\boldsymbol Z}$ recovery for ${M=20,40,63}$ using the method of \cite{chi2016guaranteed} and the proposed one. It is clear that the proposed method performs better than the analogous method. Note that the recovery of matrix ${\boldsymbol Z}$ is specifically important for estimating ${\boldsymbol x \text{ and }\boldsymbol h}$ after the spike localization. To demonstrate the performance of the proposed method in spike localization, phase transition diagrams for both methods are presented in figure \ref{figure2}. A successful recovery is considered as ${max \:\:\: |\hat{\tau}_k-\tau_k|<3\times 10^{-4},\:k=1,\ldots,K}$ where ${\hat{\tau}_k}$ is the estimation of ${\tau_k}$. As it can be seen in figure \ref{figure2}, the probability of successful spike localization is slightly higher when arbitrary sampling scheme is utilized.
\subsection{Dimension of the Subspace of PSF}\label{subDS}
\begin{figure}[!t]
  \centering
  \includegraphics[scale=0.301]{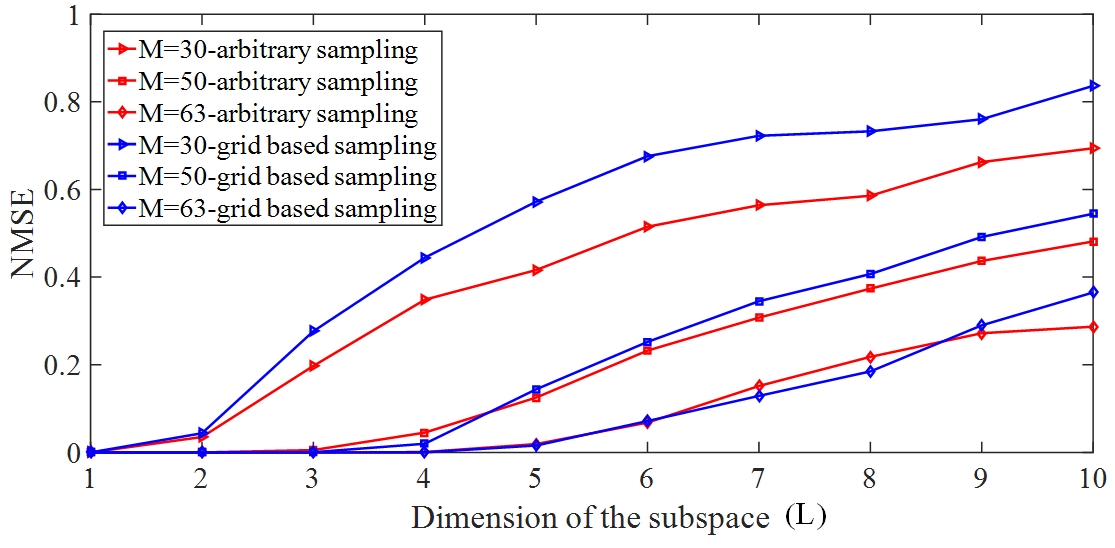}
  \caption{NMSE comparison of the proposed method (arbitrary sampling) with the method of \cite{chi2016guaranteed} for various ${L}$'s (grid-based sampling).}\label{figure3}
\end{figure}

\begin{figure}[!t]
  \centering
  \includegraphics[scale=0.155]{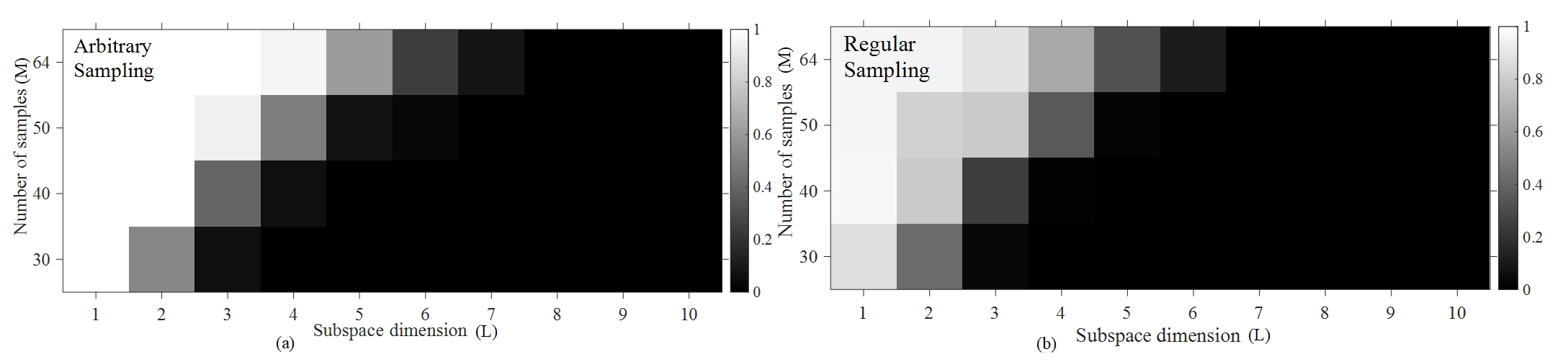}
  \caption{Phase transition diagrams for various ${L}$'s and ${M}$'s. a) The proposed method b) The method of \cite{chi2016guaranteed} }\label{figure4}
\end{figure}

In this subsection, we investigate the effect of the dimension of ${\boldsymbol h \: (L)}$ on matrix ${\boldsymbol Z}$ recovery and spike localization. In the simulations, the subspace dimension is varied from ${1\text{ to }10}$ for ${M=30,50,63}$ and ${K=5}$. Figure \ref{figure3} depicts the results. As it can be seen, the performance of the proposed method is slightly better than the method of \cite{chi2016guaranteed}. To compare the rate of successful spike localization, figure \ref{figure4} is depicted. This figure clearly suggests that the proposed method has higher success rate in spike localization than the method of \cite{chi2016guaranteed}.

\subsection{Spike Localization under Noisy Environment} \label{subSLN}
Considering more realistic scenarios, we investigated the performance of the proposed method in spike localization while the received signal is subjected to Gaussian noise ${\mathcal{N}(0,\sigma^2\boldsymbol I)}$. The SNR is defined as ${10log(\|\Upsilon(\boldsymbol Z)\|_2^2/(M\sigma^2))}$dB. Table 1 provides the MSE ${\frac{1}{N}\sum_{n=0}^{N-1}\|\hat{{\boldsymbol \tau}}_n-{\boldsymbol \tau}_n\|_2^2}$ where ${\hat{\boldsymbol \tau}_n}$ and ${{\boldsymbol \tau}_n}$ are the vectors of estimated and true spikes at ${n}$-th Monte Carlo simulation. As SNR increases, the MSE of spike localization of the proposed method decreases. Note that in noisy environments, the method of \cite{chi2016guaranteed} can not localize spikes uniquely without magnitude recovery.
  \begin{table}[t]
  \centering
  \begin{tabular}{|l|c|c|c|c|c|c|}
  \hline
  SNR & 5 & 10 & 15 & 20 & 25 & 30\\
  \hline
  MSE & 0.3262 & 0.3129 & 0.2560 & 0.1986 & 0.0941 & 0.0082\\
 \hline
 \end{tabular}
 \caption{MSE of spike localization (${K=4,L=3,M=50}$).}
 \label{table1}
 \end{table}
\section{Conclusion}
We have proposed a new method for the problem of blind super-resolution using an arbitrary sampling scheme. Incorporating arbitrary sampling scheme in the problem generalized the work of \cite{chi2016guaranteed}. More specifically, \textit{lemma }\ref{lemma1} made it possible to use arbitrary samples in the problem. After characterizing the problem in a tractable manner, a new method for the blind super-resolution problem was derived. Simulation results revealed superior performance of the proposed method than that of \cite{chi2016guaranteed}. According to the leakage of squared Fej\'{e}r kernel while using arbitrary samples, the proof of the dual certificate was left for future work.


%

\appendices
\section{}
Proof of \textit{lemma \ref{lemma1}}: Consider the following linear matrix inequality (LMI)
\begin{eqnarray}\label{12}
  \left[
    \begin{array}{cc}
      \boldsymbol W & \boldsymbol Z^{\ast} \\
      \boldsymbol Z & \boldsymbol Q \\
    \end{array}
  \right]
   &\succeq& 0.
\end{eqnarray}
For any ${\bar{\boldsymbol Q}}$ satisfying ${\boldsymbol Q=\bar{\boldsymbol Q}\bar{\boldsymbol Q^{\ast}}}$ there exists a matrix ${\bar{\boldsymbol Z}}$ so that ${\boldsymbol Z=\bar{\boldsymbol Q}\bar{\boldsymbol Z}}$. Also, ${Tr(\boldsymbol W)}$ constrained to (\ref{12}) is such that\cite{yang2016exact}
\begin{equation}\label{13}
 Tr( \boldsymbol W) \geq Tr(\boldsymbol Z^{\ast}\boldsymbol Q^{-1}\boldsymbol Z)=min\{Tr(\bar{\boldsymbol Z}^{\ast}\bar{\boldsymbol Z}):\boldsymbol Z=\bar{\boldsymbol Q}\bar{\boldsymbol Z}\}
\end{equation}

Considering ${\boldsymbol W_{\ast}}$ and ${\boldsymbol Q_{\ast}}$ as solutions to (\ref{11}), we should have ${\boldsymbol Q_{\ast}\in \mathbb{K}}$. Thus,
$${\boldsymbol Q_{\ast}=\sum_{k=1}^{P}\hat{c}_k\boldsymbol P(\hat{\tau}_k)=\sum_{k=1}^{P}[\sqrt{\hat{c}_k}\boldsymbol c(\hat{\tau_k})\boldsymbol u^{\ast}][\sqrt{\hat{c}_k}\boldsymbol c(\hat{\tau_k})\boldsymbol u^{\ast}]^{\ast}.}$$ Note that this decomposition is not necessarily unique. Using the above description, there exist complex numbers ${\{\beta_k\}_{k=1}^P}$ such that
\begin{eqnarray}\label{14}
  \boldsymbol Z &=& \sum_{k=1}^{P}\sqrt{\hat{c}_k}\boldsymbol c(\hat{\tau}_k)\beta_k\boldsymbol u^{\ast},\:\:\: Tr(\boldsymbol W_{\ast})=\sum_{k=1}^{P}|\beta_k|^2.
\end{eqnarray}
According to (\ref{13}) it is expected that ${|\beta_k|=\sqrt{\hat{c}_k}}$. Now we prove this claim. Consider ${|\beta_k|\neq\sqrt{\hat{c}_k}}$ and define$${Tr(\hat{\boldsymbol W})=\sum_{k=1}^{P}|\beta_k|\sqrt{\hat{c}_k},\hat{\boldsymbol Q}=\sum_{k=1}^{P}|\beta_k|\sqrt{\hat{c}_k}\boldsymbol c(\hat{\tau}_k)\boldsymbol c^{\ast}(\hat{\tau}_k)\in\mathbb{K}.}$$ Note that ${\|\boldsymbol u\|_2=1}$. Now it is easy to verify (\ref{12}) for ${\hat{\boldsymbol W}}$ and ${\hat{\boldsymbol Q}}$. Thus ${\hat{\boldsymbol W}}$ and ${\hat{\boldsymbol Q}}$ are in the feasible set of (\ref{10}). We have
$${Tr(\boldsymbol W_{\ast}-\hat{\boldsymbol W})+\boldsymbol e^{\ast}(\boldsymbol Q_{\ast}-\hat{\boldsymbol Q})\boldsymbol e=\sum_{k=1}^{P}\{\hat{c}_k+|\beta_k|^2-2|\beta_k|\sqrt{\hat{c}_k}\}>0,}$$ which contradicts the fact that ${\boldsymbol W_{\ast}}$ and ${\boldsymbol Q_{\ast}}$ are solutions to (\ref{10}). Thus, ${|\beta_k|=\sqrt{\hat{c}_k}}$ and ${Tr(\boldsymbol W_{\ast})=\boldsymbol e^{\ast}\boldsymbol Q_{\ast}\boldsymbol e}$ so that ${\frac{1}{2}(Tr(\boldsymbol W_{\ast})+\boldsymbol e^{\ast}\boldsymbol Q_{\ast}\boldsymbol e)=\sum_{k=1}^{P}\hat{c}_k}$. Now, (\ref{14}) is an atomic decomposition of ${\boldsymbol Z}$ since ${\boldsymbol A(\tau_k,\boldsymbol u)=\boldsymbol c(\tau_k)\boldsymbol u^{\ast}}$. According to (\ref{8}), we conclude that
\begin{equation}\label{15}
  \|\boldsymbol Z\|_{\mathcal{A}}\leq\sum_{k=1}^{P}\hat{c}_k=(Tr(\boldsymbol W_{\ast})+\boldsymbol e^{\ast}\boldsymbol Q_{\ast}\boldsymbol e)/2.
\end{equation}
To prove the other side of the inequality, we take a solution to (\ref{8}). Considering ${\tilde{P}}$ positive numbers ${\{\tilde{c}(\tau_k)\}_{k=1}^{\tilde{P}}}$ and spikes ${\{\tau\}_{k=1}^{\tilde{P}}}$ as the solution of (\ref{11}), we have
\begin{equation}\label{16}
  \|\boldsymbol Z\|_{\mathcal{A}}=\sum_{k=1}^{\tilde{P}}\tilde{c}_k,\boldsymbol Z=\sum_{k=1}^{\tilde{P}} \tilde{c}_k\boldsymbol c(\tau_k)\boldsymbol u^{\ast}
\end{equation}
By taking ${\tilde{\boldsymbol W}=\sum_{k=1}^{\tilde{P}}\tilde{c}_k\boldsymbol u\boldsymbol u^{\ast}}$ and ${\tilde{\boldsymbol Q}=\sum_{k=1}^{\tilde{P}}\tilde{c}_k \boldsymbol c(\tilde{\tau}_k)\boldsymbol c^{\ast}(\tilde{\tau}_k)}$, one can verify that ${\tilde{\boldsymbol Q}\in \mathbb{K} }$ and (\ref{12}) is positive semi-definite for ${\tilde{\boldsymbol W},\boldsymbol Z}$ and ${\tilde{\boldsymbol Q}}$. Thus,

\begin{eqnarray}
 & &  \frac{1}{2}(Tr(\boldsymbol W_{\ast})+\boldsymbol e^{\ast}\boldsymbol Q_{\ast}\boldsymbol e) = \sum_{k=1}^{P}\hat{c}_k\leq\frac{1}{2}(Tr(\tilde{\boldsymbol W})+\boldsymbol e^{\ast}\tilde{\boldsymbol Q}\boldsymbol e) \nonumber\\
    & &= \sum_{k=1}^{\tilde{P}}\tilde{c}_k=\|\boldsymbol Z\|_{\mathcal{A}}.\nonumber
\end{eqnarray}
Using (\ref{15}) and the above inequality, ${\|\boldsymbol Z\|_{\mathcal{A}}=\frac{1}{2}(Tr(\boldsymbol W_{\ast})+\boldsymbol e\boldsymbol Q_{\ast}\boldsymbol e)}$.




%
\bibliographystyle{IEEEtran}
\bibliography{references}

%








\end{document}